\documentclass[journal]{IEEEtran}
\ifCLASSINFOpdf
\else
\fi

\usepackage{cite}
\usepackage{graphics}
\usepackage{graphicx}
\usepackage{psfrag}
\usepackage{subfigure}
\usepackage{wrapfig}
\usepackage{amsmath}
\DeclareMathOperator{\Tr}{tr}
\usepackage{listings}
\usepackage{multicol}
\usepackage{relsize}
\usepackage{comment}
\usepackage{txfonts}
\usepackage{epstopdf}
\usepackage{caption}

\DeclareMathOperator*{\argmin}{argmin} 

\begin{document}
%

\title{Robust Deep Sensing Through Transfer Learning in Cognitive Radio}

\author{\IEEEauthorblockN{Qihang Peng, Andrew Gilman, Nuno Vasconcelos, Pamela C. Cosman, and Laurence B. Milstein}\\
\thanks{Qihang Peng is with the University of Electronic Science and Technology of China, Chengdu, China 611731 (email: anniepqh@uestc.edu.cn). Andrew Gilman is with the School of Natural and Computational Sciences, Massey University, Auckland, New Zealand (email: A.Gilman@massey.ac.nz). Nuno Vasconcelos, Pamela C. Cosman and Laurence B. Milstein are with the Department of Electrical and Computer Engineering, University of California at San Diego, La Jolla, CA 92093 USA (email: nuno@ucsd.edu; pcosman@ucsd.edu; lmilstein@ucsd.edu).}
}

\maketitle

\begin{abstract}
We propose a robust spectrum sensing framework based on deep learning. The received signals at the secondary user's receiver are filtered, sampled and then directly fed into a convolutional neural network. Although this deep sensing is effective when operating in the same scenario as the collected training data, the sensing performance is degraded when it is applied in a different scenario with different wireless signals and propagation. We incorporate transfer learning into the framework to improve the robustness. Results validate the effectiveness as well as the robustness of the proposed deep spectrum sensing framework.
\end{abstract}


\section{Introduction}
\label{Sec:Introduction}

Spectrum sensing enables cognitive radios to discover unused spectrum of primary users (PUs) in time, frequency and spatial domains, such that secondary users (SUs) can access these unused spectral bands to increase spectral utilization of the network \cite{REF:Zeng-2010}-\cite{REF:Peng-MILCOM2016}. Spectrum sensing is considered of critical importance for the realization of cognitive radio.

In recent years, deep learning (DL) techniques have achieved great success on many complex tasks in computer vision, speech recognition and synthesis, and natural language processing. Experience in these areas has shown that best performance is usually obtained with end-to-end models~\cite{REF:self-driving,REF:neural-machine-tranlation,REF:speech-recognition}, where a DL system learns appropriate features for the task in a data-driven fashion, instead of using engineered features, hand-crafted by domain experts. Such models may also have potential in spectrum sensing.

A DL model was proposed in \cite{REF:Lee-2017} for cooperative spectrum sensing, where the cognitive radio network (CRN) combines the individual sensing results from each SU. Measured received signal strength (RSS) or binary sensing decisions were used as the input to a deep neural network (DNN). A recent work on modulation recognition \cite{REF:O'Shea-JSTSP2018} using raw samples of the in-phase and quadrature-phase of the received temporal signals as input to a DNN shows significant gains compared to using conventional features, for example, higher order moments. However, deep learning-based approaches require significant amounts of labeled training data which follows the same distribution as the test data. In \cite{REF:GAN for sensing} and \cite{REF:GAN for attacks}, the authors propose adversarial generative networks to augment training examples, with a limited number of labeled training data, as well as domain adaptation to switch between signal types.

In this letter, we propose a DL-based spectrum sensing system, called deep sensing hereafter. 
Unlike existing DL-based spectrum sensing using expert features, the proposed method uses raw signals as 
inputs to a DNN. We observe that a DNN trained using data obtained under one set of conditions may not 
perform well when wireless conditions change, e.g., variations in wireless propagation, different PU signals. 
To improve the robustness, we propose to incorporate transfer learning \cite{REF:FineTuning}, which uses small amounts of additional data to adapt the learned models to new communications settings. Results show that transfer learning significantly improves the robustness of deep spectrum sensing.

To our knowledge, this is the first attempt at directly using signal samples rather than expert features for spectrum sensing with DL in cognitive radio, and this is the first exploration of transfer learning considering both cases of no labeled training examples and a small number of labeled training examples, toward more robust DL-based spectrum sensing. 
The rest of this letter is organized as follows. Section II presents the deep spectrum sensing algorithm and its performance. Robustness is analyzed, and two transfer learning frameworks are examined in Section III.

\section{Deep Spectrum Sensing}
\label{Sec:DeepSS}


Received radio signals pass through a rectangular bandlimited filter to limit noise, and then are sampled, producing a discrete-time sequence. A subsequence of $N$ complex-valued samples, collected during a single sensing interval, is decomposed as a $2 \times N$ real-valued vector, with the first and second row being the in-phase and quadrature components respectively, and forms a single input vector  $\textbf{x}$ to a DNN. The DNN outputs a binary class label $y$ with value $y=1$ when the PU is detected and $y=0$ when it is not.

We use a convolutional neural network (CNN) with two convolutional layers, followed by two dense layers (Table I). For the two convolutional layers, the stride is $1$ and the zero padding equals $4$. Rectified linear (ReLU) activation units are used as the non-linearity in each layer. Dropout with a rate of $0.50$ is used to regularize fully connected and convolutional layers, to reduce over-fitting. The Adam optimizer is utilized, and the last layer uses the logistic function. Given a training set of $n$ sensing interval examples $\textbf{x}_i$ and their class labels $y_i$, denoted $D=\Big\{\textbf{x}_i, y_i \Big\}_{i=1}^n$, the network parameters are learned by minimizing the empirical risk
\begin{equation}
\textbf{w}^* = \argmin_\textbf{w} \frac{1}{n} \sum_i L\big[ f(\textbf{x}_i;\textbf{w}), y_i \big]
\end{equation}
where $f(\textbf{x}_i;\textbf{w}) = p(y_i = 1 \vert \textbf{x} = \textbf{x}_i; \textbf{w})$ and the empirical risk uses the binary cross-entropy loss function
\begin{equation}
L \big[ f(\textbf{x}_i;\textbf{w}), y_i \big] =-\bigg( y_i \log\big(f(\textbf{x}_i;\textbf{w})\big) + (1-y_i) \log\big(1-f(\textbf{x}_i;\textbf{w})\big) \bigg).
\label{Eq:CrossEntroyLossFunc}
\end{equation}
\noindent This is the set of network parameters that maximizes the likelihood $\prod\nolimits_{i=1}^n f(\textbf{x}_i;\textbf{w})^{y_i}(1-f(\textbf{x}_i;\textbf{w}))^{1-y_i}$. 

The reasons for choosing a CNN are (a) relatively low complexity, (b) operation of a CNN kernel can be thought of as related to filtering operations that occur in communications receivers, and (c) the modulation recognition work by O'Shea [8] used a CNN. 
\begin{table}[!h]
\begin{center}
\caption{Deep sensing neural network} 
\begin{tabular}{|c|c|c|c|c|c|}
\hline
Layer & Output dimensions & \# of kernels & Kernel size\\ 
\hline
Input & $ 2 \times N$ & & \\
\hline
Conv1 & $256 \times 2 \times N$ & $256$ & $1 \times 9$ \\
\hline
Conv2 & $80 \times 2 \times N$ & $80$ & $1 \times 9$ \\
\hline
Dense1 & $256$ & & \\
\hline
Dense2 & $2$ & & \\
\hline
Output & $1$ & & \\
\hline
\end{tabular}
\end{center}
\end{table}

To compare the performance of spectrum sensing using deep learning, we adopt a setting where an analytical expression for the optimal sensing algorithm is available. We consider detecting a narrowband Gaussian-distributed signal in additive white Gaussian noise (AWGN), in which case the optimal sensing algorithm according to the log-likelihood ratio is\cite{REF:BookDet}
\begin{equation}
\begin{array}{llll}
LLR(\textbf{x}) = \frac{1}{2} \textbf{x}^T (C_\textbf{z}^{-1} - C_\textbf{x}^{-1})\textbf{x}
\end{array}
\label{Eq:OptSenGaussian}
\end{equation}
\noindent where $\textbf{x}$ is a vector of received samples within one sensing duration, $C_\textbf{x}$ is the covariance matrix of $\textbf{x}$, and $C_\textbf{z}$ is the covariance matrix of the additive noise after the filter.

\begin{figure}[!htb]
\centering
\includegraphics[width=.50\textwidth]{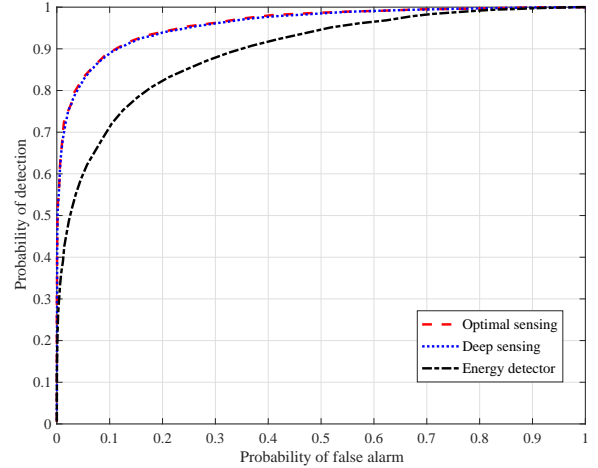}
\caption{Deep spectrum sensing compared with optimal sensing.}
\label{Fig:Opt-BandN}
\end{figure}

We compare sensing performance using a narrowband Gaussian PU signal with zero mean, corrupted by AWGN. There are $N=32$ samples in a sensing interval, and the signal-to-noise ratio (SNR) $10\log_{10}\Big(\sigma_S^2/\sigma_n^2\Big)$ is -4dB, where $\sigma_S^2$ is the PU signal variance and $\sigma_n^2$ is the noise variance after the filter. The PU signal bandwidth is $1/4$ of the filter bandwidth. The network is trained with a training set $D$ of $n = 2 \times 10^4$ and tested on an independent (but with the same transmitter, channel and receiver characteristics) test set of the same size. Fig.~\ref{Fig:Opt-BandN} shows the ROC curves for optimal and deep sensing as well as the performance of an energy detector \cite{REF:Danijela-Asilomar2004}.
The optimal sensing result was obtained with (\ref{Eq:OptSenGaussian}). The deep sensing result was obtained by computing probabilities of detection and false alarm on the test set, using different thresholds on the network output. The deep sensing, which does not require feature extraction of the received samples, 
outperforms energy detection (ED) and is close to the optimal. 

The optimal scheme for a particular sensing scenario is only optimal if it has perfect information on the required parameters. For example, the optimal scheme in Fig.~1 requires the covariance matrices of the received samples and of the additive noise after the receive filter. With estimation error in the required information, the performance degrades. Also, for different sensing scenarios, the optimal sensing scheme differs, so a dedicated sensing receiver is required for every scenario, which is costly. 

\section{Robust Deep Sensing with Transfer Learning}
\label{Sec:Transfer}

\begin{figure*}[!t]
\centering
\includegraphics[width=3.5in]{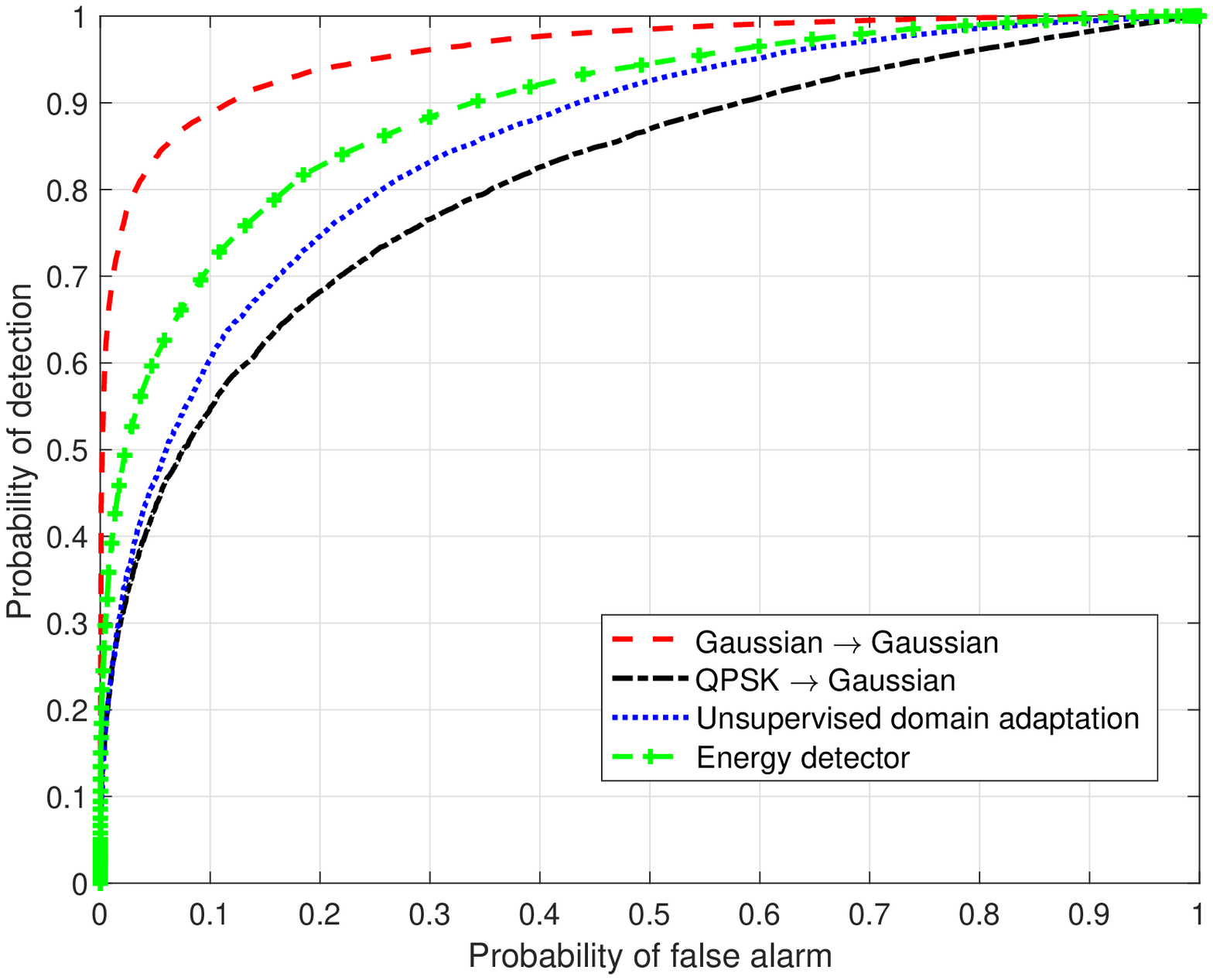}
\includegraphics[width=3.5in]{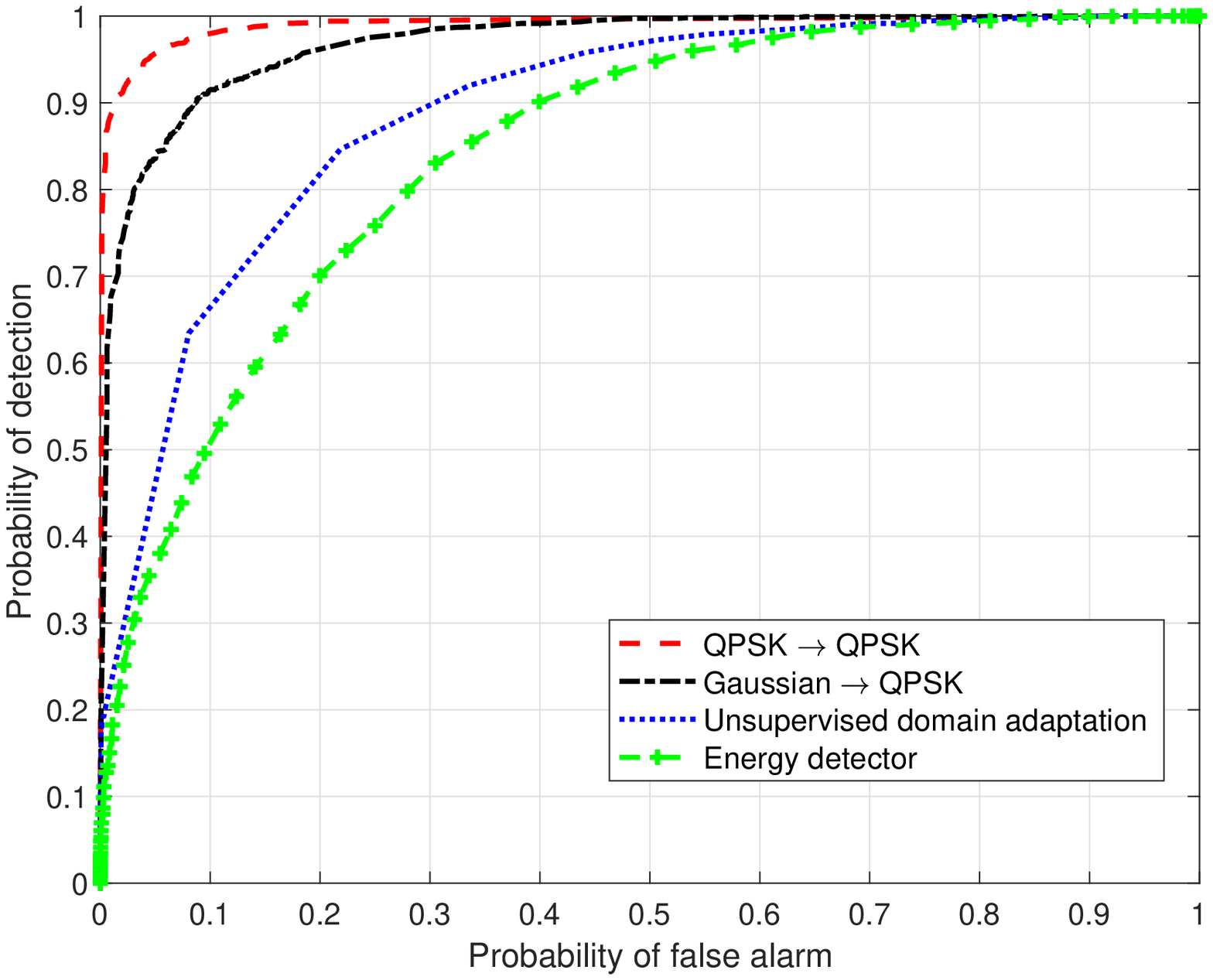}
\caption{Deep sensing using transfer learning with no labeled data: 
(left) from QPSK to zero-mean narrowband Gaussian signals; (right) from zero-mean narrowband Gaussian to QPSK signals.}
\label{Fig:TCA-NandB}
\end{figure*}

Robustness was shown to be a problem when applying DL for automatic modulation recognition 
\cite{REF:Peng-Asilomar2018}. We examine deep sensing robustness by 
considering different PU signals: narrowband Gaussian signals 
with zero mean in AWGN with an SNR of $-4\text{dB}$, and QPSK signals that use a square root raised 
cosine filter with a roll-off factor of $0.5$ as pulse shaping. 
The QPSK signals experience path loss with average SNR between $-2\text{dB}$ and $-4\text{dB}$ and 
frequency-selective Rayleigh fading with 3 discrete paths. The data is obtained from simulations in MATLAB. 
Datasets collected under these different characteristics will belong to different, but related, distributions.
We say that these datasets have been obtained in different domains. 
The {\it source} domain is used to train the network, and the {\it target} domain is used for testing.
Both training and test sets have size $n=2 \times 10^4$. 
Results are in Fig.~\ref{Fig:TCA-NandB}, where the probability of detection ($p_d$)
versus the probability of false alarm ($p_{fa}$) is plotted. 
In Fig.~\ref{Fig:TCA-NandB}(left), we use QPSK as source domain and Gaussian as target domain. 
The resulting sensing performance, marked 
``QPSK$\rightarrow$Gaussian", is significantly worse than the case where 
we use $2 \times 10^4$ examples of Gaussian signals to train and test the network 
(curve labeled ``Gaussian$\rightarrow$Gaussian"). 

Similar observations can be made from Fig.~\ref{Fig:TCA-NandB}(right), 
where the curve ``Gaussian$\rightarrow$QPSK" is 
obtained using Gaussian signals in the source domain and QPSK signals in the target domain, 
and the curve ``QPSK$\rightarrow$QPSK" is plotted for reference.
Figs.~\ref{Fig:Opt-BandN} and \ref{Fig:TCA-NandB} show that when source and target domains are the same, deep 
sensing performance can be close to optimal, whereas when they are mismatched, deep sensing performance 
can degrade significantly. 
As transmitted signals can vary in several ways (e.g., alphabet sizes, coding schemes) and 
signal propagation depends on many factors (e.g., frequency, terrain profile), 
getting enough ground-truth labeled training data across all possible scenarios is difficult.
Experience in other problems such as object recognition shows that no system is ever robust 
enough to address all possible operating conditions. Thus transfer 
learning procedures are important.

\subsection{Transfer learning with no labeled data}

The transfer approaches in this category are referred to as unsupervised domain adaptation. Let $X_{src}=\{ \textbf{x}_{src_i} \}$ and $X_{tar} = \{ \textbf{x}_{tar_i} \}$ denote the data in the source and target domains. As shown above, directly applying the neural network (NN) trained with $X_{src}$ may not work well for $X_{tar}$.
To leverage the knowledge learned by the NN from $X_{src}$, we use the transfer learning method of \cite{REF:Yang-TCA}. This aims to discover a latent space described by a kernel-induced feature transformation function $\phi$ such that the marginal distributions of $\phi\big( X_{src} \big)$ and $\phi \big( X_{tar} \big)$ are close. A nonparametric distance estimate, referred to as the Maximum Mean Discrepancy (MMD)\cite{REF:Yang-TCA}, is defined by embedding distributions in a reproducing kernel Hilbert space (RKHS) and is calculated by $\big\Vert \dfrac{1}{n_1} \sum\limits_{i=1}^{n_1} \phi(\textbf{x}_{src_i}) - \dfrac{1}{n_2} \sum\limits_{i=1}^{n_2} \phi (\textbf{x}_{tar_i}) \big\Vert_{\mathcal{H}}^2$, where $\big\Vert \cdot \big\Vert_{\mathcal{H}}$ is the RKHS norm.
Making the distributions of the source and target data close is equivalent to minimizing the MMD distance \cite{REF:Yang-TCA}. Let $\textbf{K} = \big[ \phi(\textbf{x}_i)^T \phi(\textbf{x}_j) \big]$, and $\textbf{L}_{i,j} = 1/n_1^2$ if $\textbf{x}_i, \textbf{x}_j \in X_{src}$, else $\textbf{L}_{i,j} = 1/n_2^2$ if $\textbf{x}_i, \textbf{x}_j \in X_{tar}$, otherwise, $\textbf{L}_{i,j} = -1/n_1n_2$. The MMD distance can then be written as $\Tr(\textbf{KL})$, and the learning problem formulated as \cite{REF:Yang-TCA} 
\begin{equation}
\begin{array}{llll}
&\min\limits_\textbf{W} &\Tr(\textbf{W}^T \textbf{KLKW}) + \mu\cdot \Tr(\textbf{W}^T\textbf{W})\\
& s.t. &\textbf{W}^T\textbf{KHKW} = \textbf{I}
\end{array}
\end{equation}
\noindent where $\Tr(\cdot)$ stands for the trace operation,  $\textbf{H}=\textbf{I}-(1/(n_1+n_2))\mathbf{1}\mathbf{1}^T$ is the centering matrix, $\mathbf{1}$ is a $(n_1+n_2) \times 1$ column vector with all $1$'s, a regularization term $\Tr(\textbf{W}^T\textbf{W})$ controls the complexity of $\textbf{W}$, $\mu>0$ is a tradeoff factor between the MMD distance between distributions and complexity, and $\textbf{I}$ is the identity matrix. The data in the latent space is $W^T K$, and the solution of $W$ corresponds to the $m$ $(m \leq N)$ leading eigenvectors of $(KLK+\mu I)^{-1} KHK$.

We use $p_{fa}$ and $p_d$ as the sensing performance metrics. 
Fig.~\ref{Fig:TCA-NandB}(left) shows that when QPSK data is used as source data and Gaussian data as target data, 
the transfer learning algorithm improves the sensing, 
compared to when we directly use the NN trained on QPSK data for sensing Gaussian PU signals. 
However, the improved deep sensing is still worse than ED. Further, 
interchanging source and target data, Fig.~\ref{Fig:TCA-NandB}(right) shows that unsupervised 
domain adaptation does not improve performance, although in this case either deep sensing outperforms ED. 
These results indicate that this transfer with no labeled target domain data is not robust.

\subsection{Transfer learning with a small amount of labeled data}

When we have a small amount of labeled data, we can use fine-tuning, 
the dominant transfer learning procedure in computer vision \cite{REF:FineTuning}.
The deep sensing system, trained on a large source dataset, is a starting point for further 
training using data from the target dataset. 
For training the baseline network, it is assumed that simulation data is used. 
For the transfer learning, we use simulation data also, but in practice the SU would need to acquire 
some real labeled data in its actual environment. One way to accomplish this is through cooperation 
between PUs and SUs. With a small loss of throughput, the PUs could use occasional sensing intervals for 
providing ON and OFF periods so that each SU can acquire labeled data.  Alternatively, by listening and 
comparing across consecutive sensing and data transmission intervals, an 
SU could develop estimates of the labels.

\begin{figure}[tbh]
\centering
\includegraphics[width=3.5in]{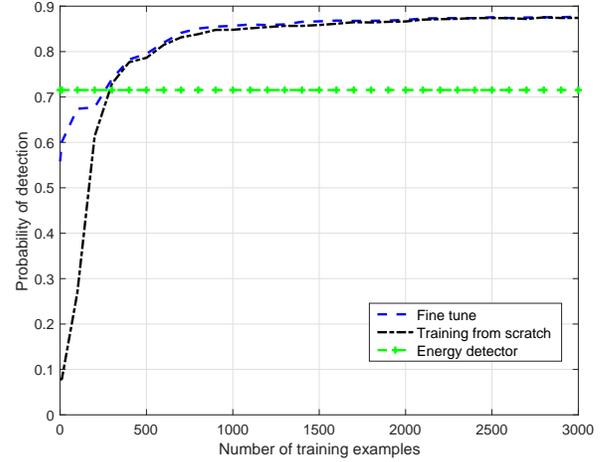}
\includegraphics[width=3.5in]{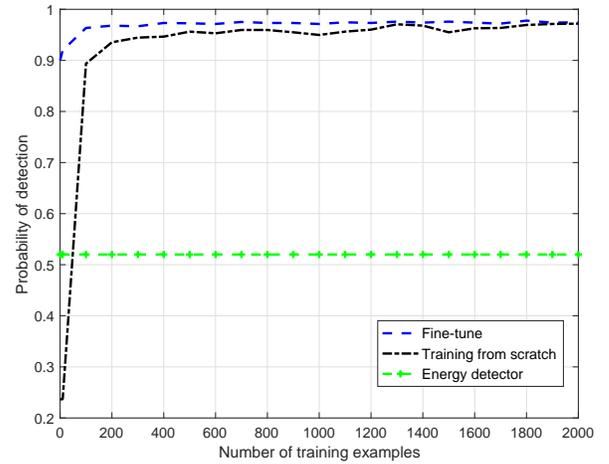}
\caption{Deep sensing performance with fine tuning: (top) from QPSK to zero-mean narrowband Gaussian signals; 
(bottom) from zero-mean narrowband Gaussian to QPSK signals.}
\label{Fig:FT-NandB}
\end{figure}

We start with a NN pre-trained using $2\times 10^4$ examples of QPSK data, and fine tune it using a variable number of examples of Gaussian signals. The fine tuned network is then applied for sensing zero-mean Gaussian signals. We also plot ED performance and the DL-based sensing performance by training from scratch which initializes the NN randomly and trains it using a variable number of Gaussian examples. To account for the stochastic nature of the stochastic gradient descent optimization with random weight initialization, the network is trained 10 times and the results are averaged. 
Fig.~\ref{Fig:FT-NandB}(top) shows $p_d$ vs.\ the number 
of examples of Gaussian signals, with $p_{fa}=0.1$. With no labeled Gaussian data, 
$p_d>0.55$ for the network trained by QPSK data, and $p_d<0.1$ for the randomly initialized network, 
showing that QPSK-trained initialization is beneficial. 
When the number of training examples is larger than roughly 300, the DL-based sensing outperforms ED. 
Fine tuning outperforms the Gaussian data training from scratch. Given enough training data, 
the performance of random initialization approaches that of the pre-trained network. 

Next we interchange the training and test data. We use Gaussian signals for pre-training and fine tune with a limited number of examples of QPSK signals.
We test by sensing QPSK signals.  As in Fig.~\ref{Fig:FT-NandB}(top), simulations are re-run 10 times 
and the results are obtained by averaging. 
Fig.~\ref{Fig:FT-NandB}(bottom) shows $p_d$ versus the number of labeled QPSK data, where  $p_{fa} = 0.1$. 
We observe a similar pattern as before: when only a small amount of QPSK training data is available, 
better performance can be achieved by fine tuning than random initialization. 
Further, fine tuning outperforms ED for the whole curve, 
and the DL-based sensing by training from scratch outperforms ED as well 
when the number of training examples exceeds roughly 100.

In addition to the narrowband Gaussian and QPSK signals, we tested
several other signals and channel models.  
For curves of the type shown in Fig.~3, the area under the curves 
over the x-axis range $\big[ 0, 1000]$ for both fine-tuning and training from scratch are in 
Table~\ref{table:results}. 
All results were consistent with Fig.~3, in that fine-tuning outperformed training from scratch. 

\begin{table}[tbh]
\begin{center}
\caption{Deep sensing performance (Area under curve) for various signals and channel models.  
In this table, PL and R denote path loss and Rayleigh fading.} 
\label{table:results}
\begin{tabular}{|c|c|c|cl}
\hline
Source domain $\rightarrow$ target domain & Fine-tune & Train from scratch\\
\hline
BPSK +PL $\rightarrow$ QPSK +PL,R & 845.64 & 673.98 \\
\hline
QPSK +PL,R $\rightarrow$ BPSK +PL & 938.72 & 849.61 \\
\hline
QPSK +PL $\rightarrow$ 16QAM +PL,R & 816.55 & 655.63 \\
\hline
16QAM +PL $\rightarrow$ BPSK +PL,R & 870.26 & 760.05 \\
\hline
\end{tabular}
\end{center}
\end{table}

{\bf Conclusion:}
We demonstrate the application of deep learning to spectrum sensing. The approach does not require 
feature extraction from the received signals at the SU. As deep spectrum sensing is not robust when 
applied in a different communications scenario from the training data, we incorporate transfer learning to 
ensure robustness.  With no labeled target data,
the transfer is unreliable and depends on whether QPSK or Gaussian signals are the source or target. 
When there is a small amount of labeled target data, 
fine tuning is shown to be robust for transferring into a variety of domains.

\end{document}